# On-Demand Teleradiology Using Smartphone Photographs as Proxies for DICOM Images

Christine Podilchuk
*Sonavista, Inc.*
Warren, New Jersey
cpodilchuk@sonavistahealth.com

Robert Warfsman
*Rutgers University*
*Professional Science Master's degrees*
New Brunswick NJ
rjw167@scarletmail.rutgers.edu

Siddhartha Pachhai
*Rutgers University*
*Professional Science Master's degrees*
New Brunswick NJ
sp1667@scarletmail.rutgers.edu

Richard Mammone
*Dept of Electrical and Computer Engineering*
*Rutgers University*
New Brunswick, NJ
rmammone@rutgers.edu

*Abstract*—The use of photographs of the screen of displayed medical images is explored to circumvent the challenges involved in transferring images between sites. The photographs can be conveniently taken with a smartphone and analyzed remotely by either human or AI experts. A denoising neural network (DnCNN) preprocessor is shown to improve the performance for human experts. The impact of the proposed method on AI performance is also evaluated and findings show the AI diagnostic accuracy to be roughly the same on any type of images but adding in the preprocessor brings additional benefit in terms both visual quality and AI diagnostic accuracy. The DnCNN preprocessor is shown to improve image quality of photographed images by 15 dB. The photo approach is an alternative to IHE-based teleradiology applications while avoiding the problems inherit in navigating the proprietary and security barriers that limit DICOM communication between PACS in practice.

## I. Introduction

Teleradiology is the transmission of radiographic images from one location to another for interpretation. Teleradiology service providers help to fill the need for sub-specialty expert consultants, vacation leaves and overflow gaps in the onsite radiology facilities A Shortage of radiologists, an increase in the use of advanced imaging methods with associated larger data files, the consolidation of hospitals into regional delivery systems, and high expectations of patients and referring physicians for timely service are among the factors that have encouraged the increasing use of teleradiology. These factors also helped to create new and potentially disruptive business models for service delivery. Radiologists have accepted the changes in organizational structure while the service expectations are taking place in the health care system that use the local and/or cloud picture archiving and communication system (PACS). Teleradiology is a large and growing industry [1].

Medical image files generally conform to the Digital Imaging and Communications in Medicine (DICOM) standard. The files reside on a picture archival and communications system (PACS). Teleradiology involves seamlessly integrating a plethora of varied PACS across locations. This scenario also involves addressing and overcoming issues related to widely varied networks, creating secure VPN or virtual private network tunnels, configuring multiple firewalls, as well as establishing and testing DICOM transfers. There is frequently a need to share data between different hospitals in a region. The problem is that different hospitals use different PACS systems and communication between PACS can be difficult.

The integration standards called the Integrating of Healthcare Enterprise (IHE) [2] have been developed to address the communication issues files between PACS. IHE is an initiative by the healthcare industry and related institutions to improve the way healthcare computer systems share information. IHE is based on a Cross-Enterprise Document Sharing profile (XDS). The advantage is images do not need to be duplicated in a central archive to be shared among the different healthcare organizations; they only need to be indexed and published in a central registry. In practice IHE standard allows different vendors and local IT departments to use different parameters in unique ways [3] which make communication between PACS very difficult if not impossible. In this paper we explore the possibility of transferring images by using a photograph of the image from the display as an alternative to transferring the DICOM image using the IHE standard. Photographs of the displayed image will differ from the original DICOM image. Hence, we explore the possibility of using a denoising CNN to transform the photographic image to an Image which is similar to the DICOM image.

## II. METHODS

The proposed photographic method is investigated for the use case of transmission of an image to an off-site human expert. For the process of transmitting photographs of medical images using smart phones, understanding the difference in quality between the transferred and original image is key in grasping the effectiveness of the system. A measure of similarity is given by the peak signal to noise ratio

$$PSNR = 20\log(255/\sqrt{MSE})$$

where MSE is the mean square error between the DICOM image and the photograph of the displayed DICOM image. Other metrics such as the structural similarity index (SSIM) have also been proposed as another perceptual metric and will be used here as well. The SSIM is calculated using the following formula

$$SSIM = \frac{(2\mu_x\mu_y + C_1)(2\sigma_{xy} + C_2)}{(\mu_x^2 + \mu_y^2 + C_1)(\sigma_x^2 + \sigma_y^2 + C_2)}$$

where all variables within the formula are mean, variances and covariances of local areas within the images being compared. In this paper we have used an algorithm that calculates the mean SSIM scores for image pairs. Some publications have shown that a PSNR of 30 dB indicates a reasonable quality for an ultrasound image. [4-6] Similarly, it has been found that an SSIM score above 0.8 is correlated with mean opinion score (MOS)[7].

To evaluate the ability of AI expert's ability in diagnosing ROI of breast cancer ultrasounds we are using area under receiver operating curve (AUROC). We will replicate the process with various transmitted data such as DICOM, photographs and DnCNN output, to get an understanding of how the AI expert is able to generalize.

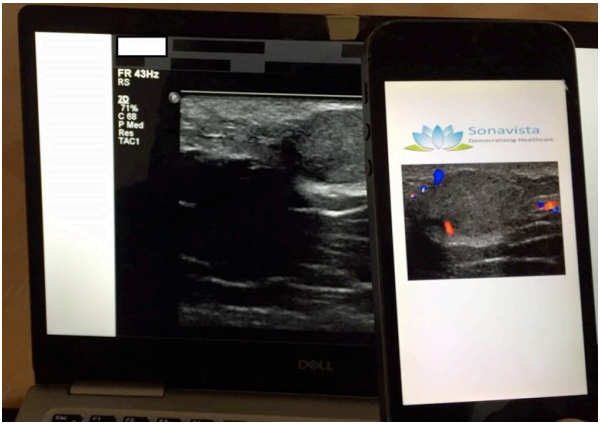

*Figure 1. Proposed teleradiology system using smartphone camera to photograph display of DICOM image providing real-time indications of disease using a CNN.*

The basic workflow structure is shown in Fig 2. The preprocessor investigated is a denoising CNN (DnCNN). The AI model used in the teleradiology experiment is an Xception CNN.

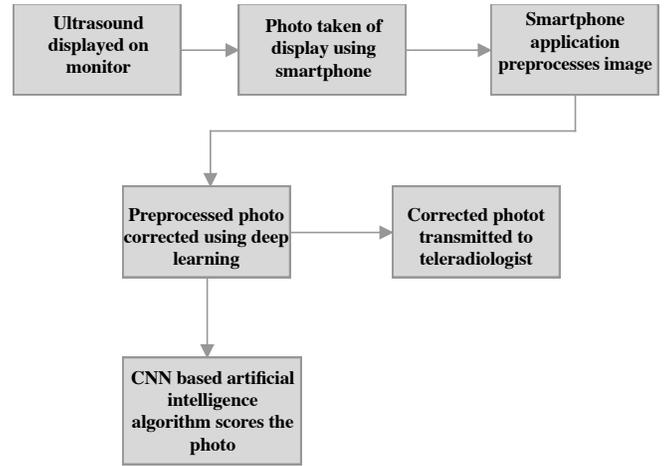

*Figure 2. Block diagram of proposed method.*

The proposed photographic method described in the workflow starts with image collection process. A total of 2862 DICOM images of breast cancer ultrasounds are used for the experiment. From out total sample, 2028 are classified benign while 834 of the images are classified malignant. For each image the region of interest containing the suspect legion is extracted and displayed on an LCD screen. A photograph is taken using a iPhone of each DICOM image from the display. The described process is envisioned to model the real-life case of a user transmitting phone captured images to a human expert for interpretation.

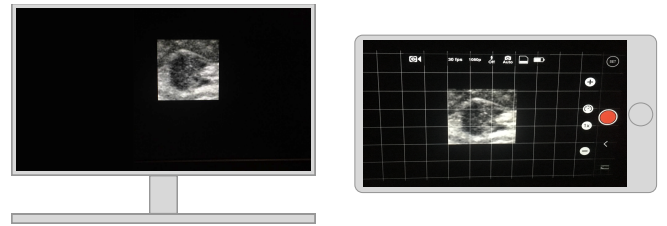

*Figure 3. ROI being displayed and a Smart Phone Being Used to Capture it.*

For the purpose of the experiment images captured from the iPhone are then adjusted to only include the ROI. The adjustment process that is used here is a perspective transformation algorithm which uses corner points as inputs in order to then flatten photographs and remove problems such as keystone. The final images are resized into the same dimension's so machine learning algorithms can be applied.

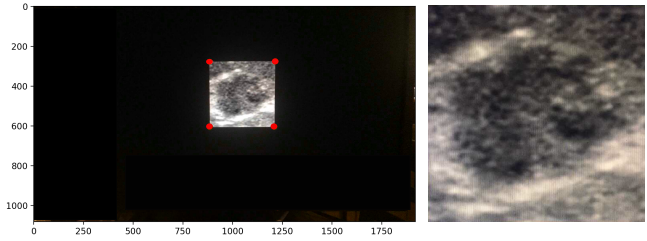

*Figure 4. Perspective Transformation applied to Photograph to extract ROI.*

PSNR & SSIM is calculated between the DICOM images and its corresponding photographic image in order to get an estimate of the difference in image quality prior to any transformation. To evaluate the effectiveness of transferring photographs we divide the data into two parts the training set which contains 2700 images and the test set which consists the remainder amount. A DnCNN is then trained using the training set with iPhone photos as input and DICOMS as the ground truth. Once the training process is completed photographs are corrected using our trained DnCNN model to map the photographs to the DICOM images, the PSNR & SSIM are recorded for the corrected images. Statistics such as average and standard deviation for the metrics PSNR & SSIM are collected for both training and test sets to evaluate the effectiveness of the photo transmission idea.

The proposed photographic method is also investigated for the scenario when the transmission of an image is to be interpreted by an AI expert. Evaluation of an AI expert is a new field which requires researchers to go through countless permutations of methodologies in order to obtain something that can be useful for real world applications. An end to end AI based solution for teleradiology that uses CNN for both image correction and later diagnosis would be an optimal solution. The collected images described above would resemble a use case where photographs are taken using an iPhone and is transmitted to an AI expert for interpretation. We also look at how the image correction described above affects the abilities of the AI expert. For the evaluation of an end to end solution we have used a cross validation approach. The 2862 images are broken into train, valid and test. The cross validation process involves ten different splits, for each split AUROC scores are recorded thrice in order to account for the variability that exists while training a neural network. For each cross-validation iteration, we record AUROC scores for the original DICOM image, photographs of the original and CNN corrected image. As the validation set is used to monitor the training process, we are only counting the AUROC scores on the test set. All training is done using the original DICOM images but for better generalization image augmentation is used in a randomized way during training.

### III. RESULTS

Our first use case where we envision a system able to transmit images through a smart phone would require the correction of color, ambient light and noise from photographs. Below is an example of what a photograph would look like. The low PSNR and SSIM evaluated is a result of the described factors affecting the iPhone photos.

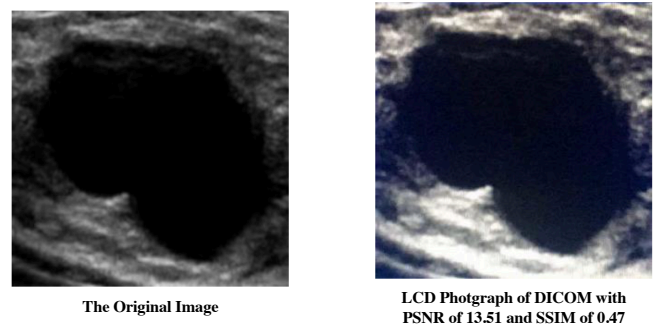

**The Original Image**     **LCD Photgraph of DICOM with PSNR of 13.51 and SSIM of 0.47**

*Figure 5*

The corrected images produced by the CNN are visually similar to the original DICOM images. The bright areas in the images captured from an LCD display have been corrected. The restoring of the image can also be observed quantitatively with a significant increase in PSNR and SSIM.

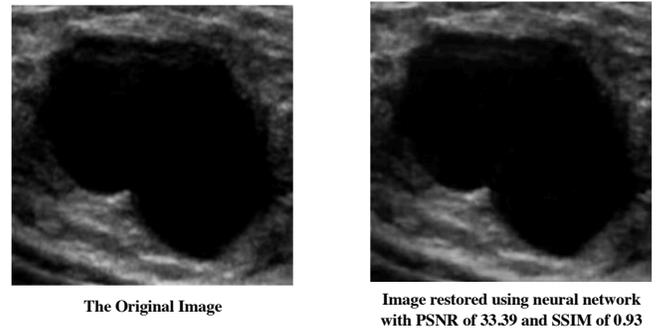

**The Original Image**     **Image restored using neural network with PSNR of 33.39 and SSIM of 0.93**

*Figure 6*

The results for the method of evaluating image quality are shown in the table below [Table 1].

*Table 1 PSNR & SSIM Results*

| Split | Count | Metric | Average Prior Transformation | Standard Deviation Prior Transofrmation | Average after Transformation | Standard Deviation after Transformation |
|---|---|---|---|---|---|---|
| Train | 2700 | PSNR | 14.5065 dB | 2.6806 dB | 30.4565 dB | 3.4801 dB |
|  |  | SSIM | 0.7118 | 0.1102 | 0.94552 | 0.02795 |
| Test | 162 | PSNR | 13.5711 dB | 1.1226 dB | 29.0058 dB | 2.6748 dB |
|  |  | SSIM | 0.704 | 0.0931 | 0.93522 | 0.02927 |

It is clearly evident that our CNN preprocessor is improving the quality of images visually and quantitively. In the use case of a medical professional receiving a photograph that is sent using a smart phone, the deep learning correction method would result in the receiver obtaining high-quality images. As described above getting a 30-dB score is considered good for a reconstructed image and in our case, the scores we are achieving are of the quality we desire. Further refining of the technique in the future will allow us to exceed these numbers. The improvement in SSIM is also proof that our scores are high enough that they would score well if MOP process was used on our images.

We have shown that image quality can be recovered using a deep learning approach, another idea that we researched was how AUROC scores were affected in the process of images being transmitted via a smart phone. We found evidence that the AI expert we had trained with data augmentation was robust to the noise and variability that were created from taking photographs. Once a photograph is taken, the average AUROC scores did drop by a small amount, the AUROC scores obtained on the original images were at 0.864 which dropped to 0.849. The use of the CNN affected the scores in a positive way, being able to lift it up to an average of 0.853 AUROC, which is closer to the original recorded score. The research indicates that CNN could be useful in mitigating the challenges that exists with developing accurate teleradiology tools.

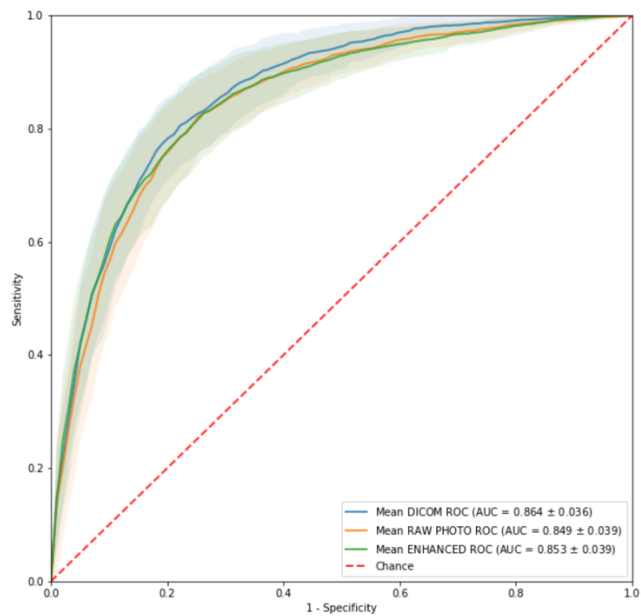

*Figure 7 AUROC Scores*

## IV Conclusion

The use of a preprocessor and DnCNN has been shown to increase the PSNR by 15 dB. The positive effects of using deep learning in the process of transferring images is not only evident in terms of image quality assurance but there seems to be good signs in terms of maintaining predictive capabilities for an AI assistant. The results demonstrate that the use of smartphone/tablet computer can be used to input medical images directly from the display screen and communicated via a separate image sharing system.

The use of photographs of the display in place of IHE/DICOM could also positively impact access and quality of healthcare services in low-and middle-income countries (LMICs) since smartphones have become ubiquitous [11] and LMICs generally have a shortage of radiologists [12]. AI models are beginning to augment/replace human expert opinions [12-15]. All findings indicate deep learning applications in teleradiology to be an exciting and useful application area.

A demonstration can be found at:
https://www.youtube.com/watch?v=ULGJY_xLwlg

## V Limitations

The current photographs were taken under ideal conditions. A forthcoming study will show the advantages of the use of a deep learning to improve photographs under nonideal conditions such as: glare, reflections, geometric distortion, color balance, greyscale errors, keystone, and motion blur